\documentclass[conference]{IEEEtran}
\ifCLASSINFOpdf
  \usepackage[pdftex]{graphicx}
   \DeclareGraphicsExtensions{.pdf,.jpeg,.png}
\else
   \usepackage[dvips]{graphicx}
   \DeclareGraphicsExtensions{.pdf, .jpg, .eps}
\fi
\usepackage[cmex10]{amsmath}
\usepackage{amssymb}
\usepackage{amsthm}
\newtheorem{prop}{Proposition}

\newtheorem{theo}{Theorem}

\usepackage{array}
\usepackage{fixltx2e}
\usepackage{url}

\begin{document}
\title{Finite Blocklength Achievable Rates for Energy Harvesting AWGN Channels with Infinite Buffer}

\author{\IEEEauthorblockN{K Gautam Shenoy, Vinod Sharma\textsuperscript{1}\footnotemark}
\IEEEauthorblockA{Dept. of ECE, Indian Institute of Science, Bangalore, India\\
Email: konchady,vinod@ece.iisc.ernet.in}}


%


\maketitle

\begin{abstract}
We consider an additive white Gaussian channel where the transmitter is powered by an energy harvesting source. For such a system, we provide a lower bound on the maximal codebook at finite code lengths that improves upon previously known bounds.
\end{abstract}

%
\IEEEpeerreviewmaketitle

\section{Introduction}
Channel capacity is the maximum rate at which one can transmit over a channel which can guarantee an arbitrarily small probability of error in transmission. However, channel capacity can be achieved as closely as required with codes of very large codeword lengths. But large codewords require more time units to transmit. Therefore it is useful to know the variation of the maximum rate at which one can transmit as a function of channel parameters and codeword length (see \cite{PPV1}).

\footnotetext[1]{Currently on sabbatical leave at Syracuse university, N. Y., USA.}

Finite blocklength results were first studied for discrete memoryless channels (DMC) by Strassen \cite{Strs} which were asymptotically tight upto the second order term. These were further refined by Polyanskiy et. al. \cite{PPV1} wherein the third order term was further refined for discrete memoryless channels and additive white Gaussian noise (AWGN) channels. Moreover in \cite{PPV2}, a general form of finite blocklength achievability bound and converse bound (also known as the meta-converse) were developed. The techniques used were based on a binary/multiple hypothesis testing model and were shown to recover several previously known achievability and converse bounds. Further improvements in the third order terms for DMC were given by Tomamichel and Tan in \cite{Tan2}. A finite-blocklength characterization for channels with state is given in \cite{Tan3}. The analysis for non-ergodic fading channels was carried out in \cite{Alle}.

The study of energy harvesting channels is motivated by wireless sensor networks (see \cite{Wsn1}). When a sensor node needs to communicate with another node, energy is expended to transmit symbols. The energy required can come from a battery or an energy harvesting system which is a system that gathers energy from some ambient source. A channel which uses an energy harvesting system has been of interest recently due to advances in wireless sensor networks \cite{Wsn2}. 

A survey on information theoretic and queuing theoretic results on energy harvesting systems may be found in \cite{Erkan}, \cite{Uluk2}. The capacity of an energy harvesting AWGN channel was obtained in \cite{VSnRaj} and \cite{Uluk1}. Finite-Blocklength achievable rate analysis for energy harvesting channels was studied for noiseless binary channels by Yang in \cite{Jyang} and for DMC and AWGN channels with infinite buffer by Fong, et. al. in \cite{Tan1}. 

If we restrict the blocklength, the achievable rates are less than the capacity of the channel. We define the backoff from capacity, as the difference between the channel capacity and the achievable rate as a function of codeword length. In \cite{Tan1}, the backoff from capacity for an energy harvesting AWGN channel with infinite buffer and harvest-use-store architecture (HUS) is shown to be $O\left(\sqrt{\frac{\log n}{n}}\right)$ where $n$ is the codeword length. In this paper, we show that a backoff of $O\left(\frac{(\log n)^a}{\sqrt{n}}\right)$ is possible for any $a>0$. Our framework is similar to that of \cite{Tan1} wherein we have an energy gathering phase where no symbols are transmitted followed by a transmission phase. The difference is that in our error analysis, we separate these two phases and bound the error terms individually as opposed to together. This allows us to use Kolmogorov's maximal inequality to upper bound the outage event leading to a simplified analysis. Moreover we are able to improve the coefficient of $\sqrt{n}$ (third order term) in the finite blocklength expression via Berry Esseen's Theorem. 

The paper is organized as follows. In Section \ref{prem}, we provide the notation and recall some known results for AWGN as well as energy harvesting AWGN channels. We also state the main theorem of this paper. In Section \ref{Proo}, we develop a detailed proof of this theorem. Section \ref{conc} concludes the paper.


\section{Preliminaries}\label{prem}
We first define AWGN channels with and without energy harvesting and provide the corresponding capacity results. We develop notation along the way.
\subsection{AWGN without energy harvesting}
An AWGN channel is characterized by an input $X_i \in \mathbb{R}$ and output $W_i \in \mathbb{R}$ at time $i$ and an input average power constraint $P$ where
\begin{equation}
W_i = X_i +\zeta_i,
\label{AWGN1}
\end{equation}
and $\{\zeta_i,~ i\geq 1\}$ are independent and identically distributed (i.i.d) normal random variables with zero mean and variance $\sigma^2$. We denote by $\mathcal{N}(\mathbf{x};A)$ the joint normal distribution with mean $\mathbf{x}$ and covariance matrix $A$. For any input $\mathbf{x} := x^{(n)} := (x_1,x_2,\cdots, x_n)$, given $n$ channel uses, the power constraint is formally given by $\sum \limits_{i=1}^n x_i^2 \leq nP$. Thus the AWGN channel can be represented by the probability density function 
\begin{equation}
\mathbf{\hat{W}}^n(\mathbf{w}|\mathbf{x})  = \mathcal{N}(\mathbf{x};\sigma^2I_n) := \frac{1}{(2\pi\sigma^2)^{n/2}} e^{\frac{-\|\mathbf{w} - \mathbf{x}\|^2}{2\sigma^2}}
\label{AWGN2}
\end{equation}
where the norm is the $L_2$ norm. Its capacity is given by
\begin{equation}
C_{G} = \frac{1}{2} \log_2\left(1 + \frac{P}{\sigma^2}\right) . \label{Cap1}
\end{equation} 
According to \cite{Strs} (also see \cite{PPV1}), the maximal code size $M(n,\varepsilon, P)$ with average error probability $\varepsilon$ for this channel is given as
\begin{equation}
\log M(n,\varepsilon,P) = nC_{G} + \sqrt{nV_G}\Phi^{-1}(\varepsilon) + O(\log n) \label{FCap1}
\end{equation} 
where $\Phi(x) = \frac{1}{\sqrt{2\pi}}\int \limits_{-\infty}^x e^{-\frac{u^2}{2}}du$ and $V_G$ is the channel dispersion parameter given by
\begin{equation*}
\frac{P(P+2\sigma^2)}{2(P+\sigma^2)^2}\log^2(e). \label{FCapV}
\end{equation*} 

In this set up, the transmitter is only subject to an average power constraint. However if we were to look at the channel with energy harvesting (EH-AWGN), the average power constraint will be replaced by the appropriate energy harvesting constraints. We expect a similar result for the EH-AWGN channel however the expressions in (\ref{FCap1}) would change.

\subsection{AWGN with energy harvesting}
\begin{figure}[ht]
	\centering
	\includegraphics[scale=0.75]{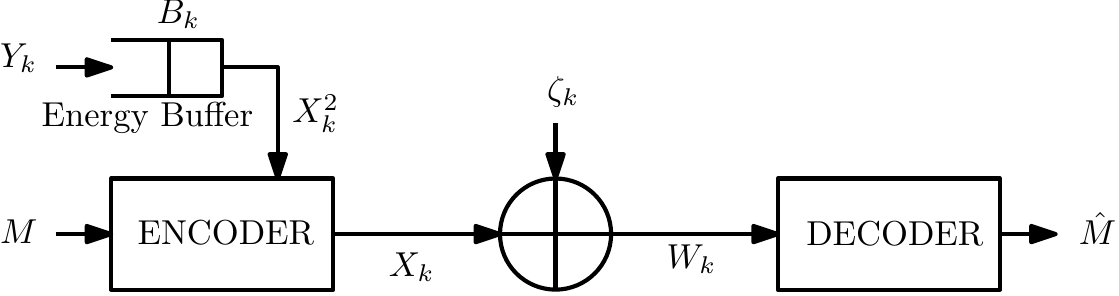}
	\caption{Block diagram of an AWGN energy harvesting system} \label{Fi1}
	\end{figure}

The harvest-use-store (HUS) model for AWGN is described in \cite{VSnRaj}. The energy harvesting system consists of a queue or buffer, which may be understood as a rechargeable battery. This buffer is recharged by an energy harvesting process. Fig. \ref{Fi1} shows the overall system setup.

In this model, there is an infinite buffer to store the harvested energy. Let $Y_i$ be the energy harvested and $B_i$ be the energy in the buffer at time $i$. Let $X_i$ be the symbol transmitted at time $i$. Then the energy used for transmission is $X_i^2$. Thus, $X_i^2 \leq B_i + Y_i$ and
\begin{equation}
B_{i+1} = B_{i} + Y_i -X_i^2.
\label{rw1}
\end{equation}
We assume $\{Y_i\}$ is i.i.d. non negative random variables with mean $E[Y]$. Let $S_n^Y = \sum \limits_{i=1}^n Y_i$. Let $Z_i = Y_i - X_i^2$ and $S_0 = 0$, $S_n = \sum \limits_{i=1}^n Z_i, ~ n\geq 1$. Thus $S_n$ is a random walk.

The capacity of the EH-AWGN channel under HUS is (\cite{VSnRaj}, \cite{Uluk1})
\begin{equation}
C_{EG} = \frac{1}{2} \log_2\left(1 + \frac{E[Y]}{\sigma^2}\right). \label{Cap2}
\end{equation} 
We hence state the main theorem of this paper.

\begin{theo}\label{Thm1}
The maximal size of the code $M^*(\hat{n},\varepsilon)$ under finite blocklength $\hat{n}$ for an EH-AWGN channel with HUS architecture, and the energy process satisfying $E[Y^2] < \infty$, satisfies
\begin{flalign}
\log M^*(\hat{n},\varepsilon) &\geq \hat{n}C_{EG} - \sqrt{\hat{n}}(\log \hat{n})^aC_{EG} \notag \\
&+\sqrt{\frac{\hat{n}V_{EG}}{2}}\Phi^{-1}\left(\varepsilon\right) -\frac{\sqrt{\hat{n}}\hat{C}}{(\log (\hat{n}/2))^{2a}} \notag \\
&-\log \hat{n} +O(1)\end{flalign}
for every $a > 0$, where $V_{EG} = \frac{E[Y]}{E[Y]+\sigma^2}$ and $\hat{C}$ is a non negative constant.
\end{theo}
Proof of this theorem is given in the next section. Comparing this result with the achievability in (\ref{FCap1}) without energy harvesting, we conclude that the price we pay in terms of bits for an energy harvesting system is of the order $O(\frac{(\log \hat{n})^a}{\sqrt{\hat{n}}})$.
\section{Proof of Theorem \ref{Thm1}} \label{Proo}

Let $0<\varepsilon<1$ and $n$ be given. We give a codebook construction that will ensure that the average probability of error is less than $\varepsilon$. The strategy follows the save and transmit strategy from \cite{Uluk1}, with some modifications.

\subsection{Codebook Generation and Encoding}
Let $M$ be the number of messages which are assumed uniformly distributed. We will eventually derive a lower bound on $M$. Generate a matrix of size $M \times n$ where each entry is generated i.i.d. with density $f_X(.)=\mathcal{N}(0; E[Y]$). Denote each row by $\mathbf{X}(m)=X^{(n)}(m), ~ 1\leq m\leq M$. This codebook is available at the decoder. 

The first $N_n$ units of time is the energy gathering phase. During this phase, the transmitter does not transmit anything but instead stores incoming energy in its buffer. Denote by $E_{0n}$ the amount of energy we expect to gather in $N_n$ time units. We shall choose $N_n$ and $E_{0n}$ towards the end of the proof. 

We denote the overall codeword length $n+N_n$ by $\hat{n}$. We count channel uses from the $N_n +1$ instant onwards. Once we gather at least $E_{0n}$ energy, we must ensure that subsequent transmissions will not empty the buffer. Let us denote the energy constraints by $\mathcal{A}_k$ where
\begin{equation*}
\mathcal{A}_k = \bigcap_{l=1}^k\{S_l \geq -E_{0n} \}.
\label{cond1}
\end{equation*} 
To send message $m$, at channel use $k$, where $1 \le k \le n$, we transmit $\hat{X}_k(m) = X_k(m)1_{\mathcal{A}_k}$. Note that the transmitted codeword satisfies the energy harvesting conditions, since $E_{0n}$ energy has already been harvested before the transmission started.


\subsection{Decoder Design}
The decoder ignores the output of the channel for the first $N_n$ channel uses as those are for harvesting energy. During the transmission stage, i.e. from the $(N_n+1)$st stage onwards, it retrieves the output of the channel $\mathbf{W} =W^{(n)}$ and then decodes the unique message $\hat{M} = m$ such that
\begin{equation}
\frac{1}{n}\log\left(\frac{\mathbf{\hat{W}}^{n}(\mathbf{W}|\mathbf{X}(m))}{\overline{\mathbf{W}}^n(\mathbf{W})}\right) > \frac{1}{n}\log(M) + \eta_n,
\label{decod1}
\end{equation}
where $\overline{\mathbf{W}}^n(.)= \mathcal{N}(0 ; (E[Y]+\sigma^2)\mathbf{I}_n)$. We choose $\eta_n >0$ later. If such a message can't be found, the decoder randomly picks one of $M$ possible messages. This decoder is known as the Threshold Decoder \cite{Tan1} (also see \cite{PPV1}).

\subsection{List of Error Events}
Denote by $\mathcal{E}$ the event that an error happens. Due to symmetry in the codebook construction, it suffices to assume that message 1 is sent. Thus,
\begin{equation*}
Pr(\mathcal{E}) = Pr(\hat{M} \neq 1| M=1).
\label{erro1}
\end{equation*}
where $Pr(A)$ denotes the probability of event $A$.

Let,
\begin{flalign*}
\mathcal{E}_{0} &= \left\{ S^Y_{N_n}  < E_{0n}\right\}, \\
	\mathcal{E}_{1} &= \bigcup_{k=1}^n \left\{ S_k < -E_{0n} \right\}, \label{outa}\\
		\mathcal{E}_2 &= \bigcup_{m=2}^M \left\{\frac{1}{n}i(\mathbf{X}(m);\mathbf{W}) > \frac{1}{n}\log(M) + \eta_n   \right\}, \\
	\mathcal{E}_3 &= \left\{\frac{1}{n}i(\mathbf{X}(1);\mathbf{W}) \leq \frac{1}{n}\log(M) + \eta_n  \right\},
\end{flalign*}
where $i(\mathbf{x};\mathbf{w}) = \log\left(\frac{\mathbf{\hat{W}}^n(\mathbf{w}|\mathbf{x})}{\overline{\mathbf{W}}^n(\mathbf{w})}\right)$. Here $\mathcal{E}_0$ is the event that insufficient energy is harvested in the gathering phase; $\mathcal{E}_1$ is the event that some part of the transmitted codeword required more energy than what was available in the buffer; $\mathcal{E}_2$ is the event that the decoder declares some other message $m \neq 1$ as the sent message and finally $\mathcal{E}_3$ is the event that the decoder failed to declare the first message as the one sent. Hence we have
\begin{equation}
Pr(\mathcal{E}) \leq \sum_{i=0}^3 Pr(\mathcal{E}_i).
\label{erro2}
\end{equation}
\subsection{Error Event $\mathcal{E}_0$}
Let $0 < E_{0n} < N_nE[Y]$. From Chebyshev's inequality, noting that $Var(Y) < \infty$, we have
\begin{flalign}
Pr(\mathcal{E}_0) &= Pr\left[ S_{N_n}^Y < E_{0n}\right] \notag\\
&= Pr\left[ S_{N_n}^Y - N_nE[Y]< E_{0n} - N_nE[Y]\right] \notag\\
&\le Pr\left[ |S_{N_n}^Y - N_nE[Y]| \geq N_nE[Y] - E_{0n}\right] \notag\\
&\le \frac{N_nVar(Y)}{(N_nE[Y] - E_{0n})^2}.
\end{flalign}

\subsection{Error Event $\mathcal{E}_1$}
We have
\begin{flalign*}
Pr(\mathcal{E}_1)  &= Pr\left[ \bigcup_{k=1}^n \left\{ S_k < -E_{0n}\right\}\right] \\
&=Pr\left[\min_{k=1,2,\cdots n} S_k < -E_{0n} \right]. \label{minny} 
\end{flalign*}
%
%
%
%
Recall that $S_n = \sum_{k=1}^nZ_k$, and $E[Z_k] = E[Y_k]-E[X_k^2] = 0$, This means $S_n$ is a zero drift random walk. Thus, by Kolmogorov's Inequality (see Chapter 3 of \cite{Athr}),
\begin{flalign*}
Pr\left(\min_{k=1,2,\cdots n} S_k < -\lambda\right) &= Pr\left(\max_{k=1,2,\cdots n} -S_k  > \lambda\right)  \\
&\leq Pr\left(\max_{k=1,2,\cdots n} |S_k|  > \lambda\right) \\
&\leq \frac{Var(S_n)}{\lambda^2}\\
&=\frac{nVar(Z_1)}{\lambda^2}.
\end{flalign*}
Since $Var(Z_1) < \infty$, picking $\lambda = E_{0n}$, we get
\begin{equation}
Pr(\mathcal{E}_1) \leq \frac{nVar(Z_1)}{E_{0n}^2}.
\label{secerrf}
\end{equation}
\subsection{Error Event $\mathcal{E}_2$}
The error probability $Pr(\mathcal{E}_2)$ may be upper bounded by a lemma proved by Shannon in \cite{Shan1} (also see \cite{Han}). We provide the proof here for completeness. Let
\begin{equation*}
\mathcal{C}_{m,n} = \left\{\frac{1}{n}i(\mathbf{X}(m);\mathbf{W}) > \frac{1}{n}\log(M) + \eta_n\right\}.
\label{cond10}
\end{equation*}
Then 
\begin{flalign}
&Pr(\mathcal{E}_2) \notag\\
&\le Pr\left(\bigcup_{m=2}^M \mathcal{C}_{m,n}\right) \notag\\
&\le \sum_{m=2}^M Pr\left( \mathcal{C}_{m,n} \right) \notag\\
&= \sum_{m=2}^M \int f_{X^n}(\mathbf{x}) \overline{\mathbf{W}}^n(\mathbf{w}) 1_{ \mathcal{C}_{m,n}}d\mathbf{x}d\mathbf{w} \notag\\
&\leq \sum_{m=2}^M \int f_{X^n}(\mathbf{x}) \mathbf{\hat{W}}^n(\mathbf{w}|\mathbf{x}) \frac{2^{-n \eta_n}}{M} d\mathbf{x}d\mathbf{w} \notag\\
&\leq 2^{-n\eta_n} \label{unionb1}
\end{flalign}
where $f_{X^n}(\mathbf{x}) = \prod \limits_{i=1}^nf_{X}(x_i)$.
\subsection{Error Event $\mathcal{E}_3$}
Let $G_i = \log \left(\frac{\mathbf{W}^1(W_i|X_i(1))}{\overline{\mathbf{W}}^1(W_i)} \right)$. Then we have
\begin{equation}
Pr(\mathcal{E}_3) \le Pr\left\{ \sum_{i=1}^n G_i \leq \log(M) + n\eta_n \right\}. 
\label{erro4}
\end{equation}

Note that $G_i$ are i.i.d. Moreover, we have
\begin{flalign*}
C_{EG} &:=E[G_i] = \frac{1}{2}\log\left(1+ \frac{E[Y]}{\sigma^2}\right), \\
V_{EG} &:=Var(G_i) = \frac{E[Y]}{E[Y] + \sigma^2}.
\end{flalign*}
Also the third moment, $E[|G_i|^3]$, is finite. To proceed further, we state the Berry Esseen's theorem (see Theorem 6.4.1 in \cite{Athr}). 
\begin{prop}[Berry Esseen's Theorem]
Let $X_i,~1\leq i\leq n$, be an i.i.d. sequence of random variables with mean $\mu$, variance $\sigma^2 <\infty$ and $E[|X_1|^3] < \infty$. Let $S_n = \sum\limits_{i=1}^n X_i$. Then we have, for any $x \in \mathbb{R}$,
\begin{equation*}
\left|Pr\left( \frac{S_n - n\mu}{\sigma\sqrt{n}} \leq x\right) - \Phi(x)\right| \leq C\frac{E|X_1 - \mu|^3}{\sigma^3\sqrt{n}},
\label{berry}
\end{equation*}
where $C<1/2$ (see \cite{Tyu}).  Note that the bound is uniform in $x$. \hfill $\square$
\end{prop}

 Let $K = \frac{E[|G_i -E[G_i]|^3]}{2V_{EG}^{3/2}}$. Applying Berry Esseen's theorem, we have for any $u \in \mathbb{R}$,
\begin{equation*}
\left|Pr\left\{ \frac{\left(\sum \limits_{i=1}^n G_i\right) -nC_{EG}}{\sqrt{nV_{EG}}} \leq u \right\} - \Phi(u)\right| \leq \frac{K}{\sqrt{n}}.
\label{bery1}
\end{equation*}
Substituting $u =  \frac{\log M + n(\eta_n - C_{EG})}{\sqrt{nV_{EG}}}$, we get

\begin{flalign}
&Pr\left\{ \sum_{i=1}^n G_i \leq \log(M) + n\eta_n \notag \right\} \\
&\leq \Phi \left(\frac{ \log M+n( \eta_n - C_{EG}) }{\sqrt{nV_{EG}}}\right) + \frac{K}{\sqrt{n}}. \label{bery2}
\end{flalign}

Let 
\begin{align}
\varepsilon_n &= \varepsilon - \frac{N_nVar(Y)}{(N_nE[Y] - E_{0n})^2} - \frac{nVar(Z_1)}{E_{0n}^2} \notag \\
&\quad- 2^{-n\eta_n} -\frac{K}{\sqrt{n}}.\label{errorq} 
\end{align}
We will specify the different constants $N_n$, $E_{0n}$ and $\eta_n$ such that by picking $n$ large enough, we get $\varepsilon_n > 0$. Taking

\begin{equation}
\log M \leq nC_{EG} + \sqrt{nV_{EG}}\Phi^{-1}\left(\varepsilon_n\right)  -n\eta_n
\label{bery3}
\end{equation}
for $n$ large enough, will ensure that the RHS of (\ref{bery2}) is less than $\varepsilon_n +\frac{K}{\sqrt{n}} $. This indicates that as long as $M$ satisfies the above equation, the average probability of error can be made less than or equal to $\varepsilon$. Since the optimum code size can only be larger than the RHS of (\ref{bery3}), we conclude that the largest code size satisfies

\begin{flalign}
\log M^*(n+N_n,\varepsilon) \geq nC_{EG} + \sqrt{nV_{EG}}\Phi^{-1}\left(\varepsilon_n\right) -n\eta_n - 1. \label{bery4}
\end{flalign}

We further simplify $\Phi^{-1}\left(\varepsilon_n\right)$ using Taylor's theorem. There exists $u \in (\varepsilon_n , \varepsilon)$ such that

\begin{equation*}
f\left(\varepsilon_n\right) = f\left(\varepsilon\right) +(\varepsilon_n-\varepsilon)f'(u),
\label{tayl2}
\end{equation*}
where $f(x) = \Phi^{-1}(x)$. Note that $f(x)$ has a derivative that is positive, strictly decreasing upto $x=1/2$; beyond which it increases. Thus in $(\varepsilon_n, \varepsilon)$, $f'(u) \leq \hat{f} = \max\{f(\varepsilon_{n_0}), f(\varepsilon)\}$ where $n_0$ is the smallest $n$ for which $\varepsilon_n > 0$. Hence we get

\begin{flalign*}
\log M^*(n+N_n,\varepsilon) &\geq n(C_{EG}-\eta_n) + \sqrt{nV_{EG}}\Phi^{-1}\left(\varepsilon\right) \notag\\
&+  \sqrt{nV_{EG}}(\varepsilon_n -\varepsilon)\hat{f} -1.
\label{bou1}
\end{flalign*}

We pick $N_n$,  $E_{0n}$ and $\eta_n$ such that
\begin{enumerate}
	\item $N_n = O(\sqrt{n}(\log n)^a)$ for $a>0$. 
	\item $0 < E_{0n} < N_n(E[Y])$.
	\item $\eta_n = o(\sqrt{n})$ and $\varepsilon_n - \varepsilon = o(\sqrt{n})$. 
\end{enumerate}
This ensures that the upper bounds for $Pr(\mathcal{E}_0)$, $Pr(\mathcal{E}_1)$ and $Pr(\mathcal{E}_2)$ tend to zero as $n \to \infty$. In the following, we set $\eta_n = \frac{\log(n)}{n}$, $N_n = \sqrt{n}(\log n)^a$ and $E_{0n} = \frac{N_nE[Y]}{2}$. Thus,

\begin{flalign}
\sqrt{n}(\varepsilon_n - \varepsilon) &= -\frac{4 Var(Y)}{(\log n)^a(E[Y])^2} - \frac{4 \sqrt{n}Var(Z)}{(E[Y])^2(\log n)^{2a}} \notag \\
&- \frac{1}{\sqrt{n}} - K.
\end{flalign}
Gathering all terms together, we get 

\begin{flalign}
\log M^*(n+N_n,\varepsilon) &\geq nC_{EG} + \sqrt{nV_{EG}}\Phi^{-1}\left(\varepsilon\right) \notag\\
&-\frac{4\sqrt{nV_{EG}}Var(Z)\hat{f}}{E[Y]^2(\log n)^{2a}} -\log n \notag \\
&+ O(1).
\label{bou2}
\end{flalign}
Substituting $\hat{n} = n+N_n$,

\begin{flalign}
\log M^*(\hat{n},\varepsilon) &\geq (\hat{n} -N_n)C_{EG} + \sqrt{(\hat{n}-N_n)V_{EG}}\Phi^{-1}\left(\varepsilon\right) \notag\\
&-\frac{4\sqrt{(\hat{n}-N_n)V_{EG}}Var(Z)\hat{f}}{E[Y]^2[\log(\hat{n} - N_n)]^{2a}} -\log(\hat{n} -N_n) \notag \\
&+O(1) \notag \\
&\ge \hat{n}C_{EG} -N_nC_{EG} + \sqrt{(\hat{n}-N_n)V_{EG}}\Phi^{-1}\left(\varepsilon\right) \notag \\
&-\frac{4\sqrt{\hat{n}V_{EG}}Var(Z)\hat{f}}{E[Y]^2[\log(\hat{n} - N_n)]^{2a}} -\log(\hat{n}) +O(1). \notag \label{bou3}
\end{flalign}
We note that for $n$ large enough, $\frac{N_n}{\hat{n}} < \frac{1}{2}$ and hence,
\begin{equation*}
\sqrt{(\hat{n}-N_n)V_{EG}}\Phi^{-1}\left(\varepsilon\right) \geq \left\{\begin{array}{l l}
\sqrt{\hat{n}V_{EG}}\Phi^{-1}\left(\varepsilon\right) & 0 < \varepsilon < \frac{1}{2}, \\
\sqrt{\frac{\hat{n}}{2}V_{EG}}\Phi^{-1}\left(\varepsilon\right) & \frac{1}{2} < \varepsilon < 1, 
\end{array}\right.
\label{bou4}
\end{equation*}
Thus we get for every $0<\varepsilon <1$, and $\hat{n}$ large enough

\begin{flalign}
\log M^*(\hat{n},\varepsilon) &\geq \hat{n}C_{EG} - \sqrt{\hat{n}}(\log \hat{n})^aC_{EG} \notag \\
&+\sqrt{\frac{\hat{n}V_{EG}}{2}}\Phi^{-1}\left(\varepsilon\right) -\frac{\sqrt{\hat{n}}\hat{C}}{(\log (\hat{n}/2))^{2a}} \notag \\
&-\log \hat{n} +O(1),
\label{bou5}
\end{flalign}
for some constant $\hat{C}$ which concludes this proof. Moreover dividing by $\hat{n}$ gives us 
\begin{equation*}
\frac{\log M^*(\hat{n},\varepsilon)}{\hat{n}} \geq C_{EG} - O\left(\frac{(\log\hat{n})^a}{\sqrt{\hat{n}}}\right).
\label{bou6}
\end{equation*}
Thus, we see that the backoff from capacity is $O\left(\frac{(\log \hat{n})^a}{\sqrt{\hat{n}}}\right)$.

\section{Concluding Remarks}\label{conc}
We have shown that for energy harvesting AWGN channels under HUS, it is possible to achieve a backoff of $O\left(\frac{(\log \hat{n})^a}{\sqrt{\hat{n}}}\right)$. Moreover we have improved the coefficient of $\sqrt{\hat{n}}$ which for $\varepsilon > 1/2$ will positively contribute to the finite blocklength bound. It is not known if the backoff can be improved to $O(1/\sqrt{\hat{n}})$ but, as is evident from the proof technique, the key lies in bounding $Pr(\mathcal{E}_1)$ by a stronger inequality than Kolmogorov's inequality. 


\section*{Acknowledgment}

The authors would like to thank Himanshu Tyagi and Deekshith P. K. for their valuable inputs and helpful discussions.



%

\end{document}